\documentclass[pre,showpacs,twocolumn,superscriptaddress,floatfix]{revtex4}
\usepackage{dcolumn,graphicx,amsmath,amssymb,txfonts}

\newcommand{\comment}[1]{}

\begin{document}

\title{A dual assortative measure of community structure}
\author{Todd D. Kaplan} \affiliation{Department of Computer Science,
  University of New Mexico, Albuquerque, NM 87131, U.S.A.}

\author{Stephanie Forrest} 
\affiliation{Department of Computer
  Science, University of New Mexico, Albuquerque, NM 87131, U.S.A.}
\affiliation{Santa Fe Institute, 1399 Hyde Park Road, Santa Fe, NM
  87501, U.S.A.}

\begin{abstract}
  Current community detection algorithms operate by optimizing a
  statistic called \emph{modularity}, which analyzes the distribution
  of positively weighted edges in a network. Modularity does not
  account for negatively weighted edges. This paper introduces a dual
  assortative modularity measure (DAMM) that incorporates both
  positively and negatively weighted edges. We describe the the DAMM
  statistic and illustrate its utility in a community detection
  algorithm. We evaluate the efficacy of the algorithm on both
  computer generated and real-world networks, showing that DAMM
  broadens the domain of networks that can be analyzed by community
  detection algorithms.
\end{abstract}
\pacs{89.75.Fb, 89.75.Hc}

\maketitle

The problem of detecting community structure within complex networks
has received considerable attention in recent
literature~\cite{newman:finding}~\cite{newman:fast}~\cite{newman:spectral}~\cite{newman:spectral2}~\cite{arenas}~\cite{clauset:large}.
Given a network of nodes and edges, the challenge is to group nodes
into communities according to the distribution of edges. There exist
many possible ways to define \emph{community} mathematically.  One
widely accepted definition, known as
\emph{modularity}~\cite{newman:finding}, defines a community to be a
group of nodes that are more densely connected than would be expected
if the edges had been assigned at random. This definition assumes that
each edge has a positive weight.

\begin{figure}
  \resizebox*{0.9
    \linewidth}{!}{\includegraphics[angle=-90]{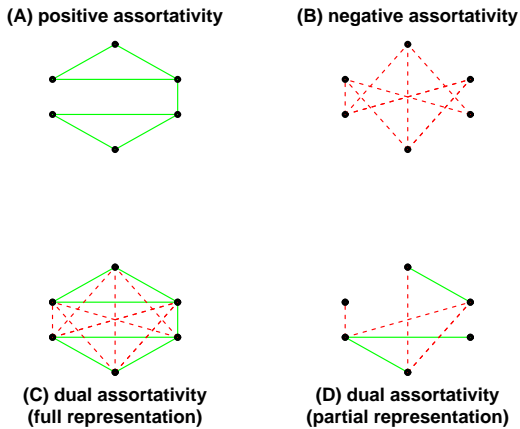}}
  \caption{Comparison of networks with different types of
    assortativity. Solid edges denote positive edge weights; dashed
    edges denote negative edge weights. The network in panel A
    portrays positive assortativity (PA) and provides an example of a
    friendship network. In panel B, the network strictly contains
    negative edges and depicts negative assortativity (NA).  We refer
    to this type of network as an adversarial network. The network in
    panel C exemplifies a fully connected dual assortativity (DA)
    network. Here, both positive and negative edges are present. In
    panel D, a partially connected dual assortative network is
    illustrated. The dual assortative modularity measure (DAMM) can be
    used to assess community structure in all of the above cases.  The
    partially connected DA network of the bottom right is the most
    general, and it is these types of networks that we study in
    section \ref{sec:results}.}
  \label{fig:assortative-networks}
\end{figure}

A common example of such a network is a \emph{friendship network}.
Nodes of the friendship network represent people, and the edges, which
are positively weighted, represent friendships.  Intuitively,
communities are comprised of sub-graphs in the network that are
densely connected to one another but sparsely connected to the
outside.
The term \emph{assortativity} ~\cite{newman:assortative}
~\cite{newman:mixing} refers to the tendency for nodes to be connected
to others that are like, or unlike, them. In the case of the
friendship network, communities are based on \emph{positive
assortativity} because nodes are connected to others with whom they
share a positive connection (friendship).  Panel A of figure
\ref{fig:assortative-networks} depicts a friendship network, where the
solid edges are positively weighted and represent friendships.

In this paper, we incorporate the concept of \emph{negative
  assortativity}, or disassortativity, into the definition of
community.  Nodes are negatively assortative if their connection is
based on dissimilarity, rather than likeness. With regards to the
friendship network, negatively weighted edges represent the strength
of adversarial relationships. We refer to a network that contains only
negatively weighted edges as an \emph{adversarial network}. As
previously described, all of the edges in the network shown in panel A
of figure \ref{fig:assortative-networks} are based on friendships.
However, let us assume that this friendship information is unavailable
and that instead a list of adversarial relationships between nodes is
provided. Further, assume that all pairings in the original network
that did not share friendships are now considered adversaries. The
resulting adversarial network is presented in panel B of figure
\ref{fig:assortative-networks}. The dashed edges indicate negative
weights. The two networks, the friendship network (top left) and the
adversarial network (top right), provide similar but different
information. It is not the case that the adversarial network is always
the reciprocal of the friendship network.

In this paper, we combine the two concepts -- both positive and
negative assortativity -- to form a single definition of
community. Because this definition incorporates the contributions of
both negative and positive relationships, we refer to it as \emph{dual
assortative}. The networks on the bottom of figure
\ref{fig:assortative-networks} illustrate this duality. They contain
both positive relationships (solid edges) and negative relationships
(dashed edges). Such networks may be fully connected, as is the case
of the network in panel C of figure
\ref{fig:assortative-networks}. However, more commonly, only a
fraction of the possible relationships between nodes may be known
(panel D of figure \ref{fig:assortative-networks}). An example of a
dual assortative network is one in which the edge weights are based on
a similarity measure, such as correlation, which can assume either a
positive or negative value. Consider a network where the nodes
represent financial traders and the edge weights indicate the
correlation of trading behavior between a pair of traders. The dual
assortative modularity measure (DAMM) definition incorporates all
available information, positive and negative, to assess the strength of
community structure.

Intuitively, there exists an asymmetry in the information provided by
positive and negative edges. A friendship between two people conveys a
stronger bond than sharing a common adversary. However, this is only
true when there are three or more communities. When only two
\emph{real} communities exist, a negative edge provides the same
amount of information as a positive edge.  Consider the case of having
two communities, community $C_a$ and community $C_b$, if two nodes
share a negative edge it indicates that one node should belong to
$C_a$ and the other to $C_b$. However, in the case of three or more
communities, the negative edge simply indicates that the nodes should
reside in separate communities but does not indicate which particular
communities the nodes should belong. The information provided by a
positive edge is more specific than that for a negative edge. 

The remainder of the paper is organized as follows. In section
\ref{sec:methods}, the mathematical framework for the dual assortative
measure is introduced and explained. Further, a community detection
algorithm used for optimizing the DAMM is described. In section
\ref{sec:results}, we assess the efficacy of optimizing the DAMM on
both computer generated and real networks, and section
\ref{sec:conclusion} summarizes and concludes the paper.

\section{Methods}
\label{sec:methods}
This section introduces the mathematics of the dual assortative
modularity measure (DAMM), describes the extremal optimization
algorithm for optimizing the DAMM on a given network, and describes a
measure, called communal overlap, which we use to quantify the
fidelity of a community detection result given that the real
communities are known and available for comparison.  

\subsection{The dual assortative modularity measure}
Before describing the dual assortative modularity measure (DAMM), we
review the original modularity measure, which provides the foundation
for the DAMM.  We then show how to quantify the negative assortative
contributions of a network. The positive and negative components are
then combined to establish the DAMM.  Finally, we introduce the
algorithm used to optimize the DAMM on a network.

\subsubsection{Positive assortativity}
\label{sec:pos-assortative}
We denote an edge weight between node $r$ and node $s$ as $e_{rs}$.
For simplicity of explanation, we consider only networks with edges of
weight $e_{rs} \in \{-1,0,1\}$, with $e_{rs} = 0$ implying that the
edge is not present.  However, in
practice $e_{rs}$ will often be a real number, $e_{rs} \in
\Re$. Given a set of communities, denoted ${C}$, we use 
$\{w_{ij} = \sum_{r s} e_{rs} | r \in C_i, s \in C_j\}$ to denote the
cumulative edge weight between community $i$ and community $j$. 

Equation \ref{eqn:mod-pos} gives the original modularity
measure~\cite{newman:finding}, denoted as $Q^+$. The implied edge
weight domain is $e_{rs} \in \{0,1\}$, which is analogous to an
\emph{unweighted} network.  The $w_{ii}$ term represents twice the
number of edges in which both ends terminate at nodes belonging to
community $i$.  Further $a_i = \sum_{j} w_{ij}$ gives the sum of all
edge weights with at least one end attached to a node residing in
community $i$, and $T = \sum_i \sum_{j} w_{ij}$ is twice the total
number of edges in the network. We use the term \emph{spoke} to refer
to the terminal end of an edge. With reference to $a_i$, we count the
number of spokes connected to community $C_i$. Similarly, $T$ refers
to the total number of spokes in the network.

\begin{eqnarray}
  \label{eqn:mod-pos}
  Q^+ = \sum_{i=0}^{|{C}|-1} \frac{w_{ii}}{T} - \left( \frac{a_i}{T}\right)^2
\end{eqnarray}

The DAMM is given in equation \ref{eqn:mod-dual}.  It uses a modified
form of equation \ref{eqn:mod-pos} to compute the contribution of
positively weighted edges.  Specifically, we redefine $a_i = \sum_{j}
H(w_{ij}) w_{ij}$, where $H(x)$ is the unit step function.  This
modification ensures that $a_i$ incorporates only the contributions of
positively weighted edges. We also redefine $T$ as $T = \sum_i
\sum_{j} |w_{ij}|$, so that both positively and negatively weighted
edges contribute to the total weight $T$.

The summation in equation~\ref{eqn:mod-pos} iterates through the set of
communities.  For each community, the difference
$\left(w_{ii}/T\right) - \left(a_i/T\right)^2$ reflects the strength
of that particular community.  The first term,
$\left(w_{ii}/T\right)$, represents the ratio of intra-communal edges
to the total number of edges in the network. An edge is considered to
be intra-communal if both ends are connected to nodes residing in the
same community. One could mistakenly assume that the higher this
ratio, the greater the strength of the community.  However, if it
were, the ratio would be optimized by a single community containing
all nodes within the network.  Thus, we compare the ratio found in the
first term to the expectation of its value, $\left(a_i/T\right)^2$.
The ratio $a_i/T$ represents the ratio of edge spokes connected to the
given community to the total number of edge spokes, and thus its square
gives the expectation.
If the difference is positive, the observed ratio is greater than what
would be expected if the edges were placed randomly.  The greater the
(positive) difference, the greater the communal strength.  If the
difference is negative, the communal strength is found to be weaker
than the expectation, suggesting no communal structure for the
community being investigated.

\subsubsection{Negative assortativity}
\label{sec:neg-assortative}

This measure is motivated by the idea that a shared adversary represents a
commonality. In other words, if both Jack and Jill are both adversaries
with Alice, they share a commonality regardless of whether the pair
are friends. 
In the case of positive weights, we quantified how much the ratio of
edges encapsulated by a community differed from the expected value
under random edge assignments. Here,
the scenario is reversed.
Adversarial relationships within a community are not desirable. In a
scenario of perfect community structure, all negative edges would
occur between communities.

Equation \ref{eqn:mod-neg} defines the negative assortative component
of the DAMM. The equation resembles that of equation
\ref{eqn:mod-pos}, except the order of terms is reversed and we
consider negatively weighted edges.  Here, $\bar{a_i}$ represents the
cumulative negative edge weight connected to nodes of community $i$.
In other words, $\bar{a_i} = \sum_j [1-H(w_{ij})] w_{ij}$, where
$H(x)$ is the unit step function. $\bar{w_{ii}}$ represents the
cumulative negative edge weight encapsulated by community $i$. The
first term of equation \ref{eqn:mod-neg} provides a null test.

\begin{eqnarray}
  \label{eqn:mod-neg}
  Q^- = \sum_{i=0}^{|{C}|-1} \left(\frac{\bar{a_i}}{T}\right)^2 - 
\frac{\bar{w_{ii}}}{T} 
\end{eqnarray}

\subsubsection{Dual assortative modularity measure (DAMM)}
To establish the DAMM, we combine the negative assortativity
contribution of equation \ref{eqn:mod-neg} with the positive
assortativity contribution of equation \ref{eqn:mod-pos}. We define
the DAMM as follows:

\begin{eqnarray}
  Q^D &=& Q^+ + Q^-\\
  &=& \sum_{i=0}^{|{C}|-1} \left[\frac{w_{ii}}{T} - 
    \left(\frac{a_i}{T}\right)^2\right] - 
  \sum_{i=0}^{|{C}|-1} \left[\left(\frac{\bar{a_i}}{T}\right)^2 -
    \frac{\bar{w_{ii}}}{T}\right] \\ 
  &=& \sum_{i=0}^{|{C}|-1} \left(\frac{w_{ii} - \bar{w_{ii}}}{T}\right) +
  \left(\frac{\bar{a_i}^2 - a_i^2}{T^2}\right)
  \label{eqn:mod-dual}
\end{eqnarray}

In the absence of negative edges, $\bar{w_{ii}}=0$ and $\bar{a_i}=0$
for all $i$, and the DAMM reduces to equation \ref{eqn:mod-pos}.

\subsection{Optimizing DAMM with the extremal optimization algorithm}
We use extremal optimization
(EO)~\cite{boettcher:eo-prl}~\cite{boettcher:eo-pre} to optimize the
DAMM statistic and detect communities in a network. EO is known to be
effective for community detection using the original modularity
measure that is based solely on positive assortativity~\cite{arenas}.
EO is a divisive approach, in which all nodes are initially placed in
a single community. Thereafter, each community is divided recursively
into two independent communities, not necessarily of the same size.
At each step, the division found to provide the largest increase in
modularity is applied, given that the increase is positive. If the
best division does not increase modularity, the community is declared
\emph{indivisible}.  When all existing communities are found to be
indivisible, the algorithm halts.

Each division proceeds as follows. Initially, the nodes are randomly
assigned to one of two partitions. After all nodes have been assigned,
the DAMM is computed. Thereafter, a single node is migrated from one
partition to the other, and the DAMM is recomputed by adding $\Delta Q^D$
associated with the migrated node (section \ref{sec:damm-derivative}).

A counter, denoted $K$, tracks the number of moves since the last DAMM
improvement. If the DAMM fails to improve, the counter is
incremented. Otherwise, the counter is reset to zero, and the
partitioning is recorded along with its associated DAMM value.  This
partitioning represents the best detected configuration. The process
continues until the counter reaches a predetermined threshold.  For
each division, the size of the community, denoted $|C_i|$, determines
the stopping criterion such that the maximum allowable number of steps
is $S=\alpha |C_i|$ ($\alpha=3$ in the experiments of section
\ref{sec:results}).  Once the counter reaches the threshold $S$, such
that $K=S$, the process terminates and the best detected configuration
is retained. If the split has improved the DAMM value, the global set
of communities is updated to reflect the division.  Otherwise, it
remains unchanged and is marked indivisible.

\subsubsection{Calculating $\Delta Q^D$}

\label{sec:damm-derivative}
An important component of the EO algorithm involves choosing which
nodes to migrate. Rather than choosing nodes at random, we associate a
value $\Delta Q^{D,u}$ with each particular node $u$.  This approach
differs slightly from that of ~\cite{arenas}, which uses a heuristic
to $\Delta Q$ rather than the exact difference. The value $\Delta
Q^{D,u}$ represents the change in DAMM that occurs by migrating the
specified node. This method resembles hill-climbing used in other
settings and biases the search for an optimal division towards
immediate improvements.

In practice, we maintain a list that associates a
$\Delta Q^{D,u}$ value with each node $u$.  To select a node for
migration, we rank the list of $\Delta Q^{D,u}$ values and then
probabilistically choose a node using a method known as
$\tau$-EO~\cite{arenas}~\cite{boettcher:eo-prl}. Using this process, a
node of rank $q$ is chosen with probability of $P(q) \approx q^{-\tau}$
where $\tau = 1 + \frac{1}{\log |C_i|}$. Following the migration of a
node $u$, $\Delta Q^{D,u}$ is updated as described in section
\ref{sec:damm-second-derivative}.

The calculation of $\Delta Q^{D,u}$ is given by equation \ref{eqn:qDamm}.
The derivation is provided in appendix \ref{appdx:first-derivative}.

\begin{eqnarray}
  \Delta Q^{D,u}_t &=& \Delta Q^{+,u}_t + \Delta Q^{-,u}_t\\
  &=& 2 \left[\left(\frac{w_{gu}-w_{lu}}{T}\right) + \frac{d_u}{T^2}
    \left(a_l-a_g-d_u)\right) \right] \nonumber \\
  && + 2 \left[\left(\frac{\bar{w}_{lu}-\bar{w}_{gu}}{T}\right)
   + \frac{\bar{d}_u}{T^2}
   \left(\bar{a}_g-\bar{a}_l+\bar{d}_u)\right) \right]
  \label{eqn:qDamm}
\end{eqnarray}

\subsubsection{Calculating $\Delta^2 Q^D$}
\label{sec:damm-second-derivative}
After each migration, all $\Delta Q^{D,u}$ values are subject to
change.  Rather than recompute the value for each node, a less
computationally expensive approach makes use of $\Delta^2 Q^{D,um}$.
Each value can be updated as $\Delta Q^{D,u}_{t+1} = \Delta Q^{D,u}_t +
\Delta^2 Q^{D,um}_t$, where $\Delta^2 Q^{D,um}_{t}$ represents
represents the change in $\Delta Q^{D,u}$ for node $u$ following the
migration of node $m$ at time $t$.

As noted, computation of $\Delta^2 Q^{D,um}_t$ involves two nodes: the
node $m$ that was migrated and the node $u$ for which $\Delta Q^{D,u}$
must be updated. Both nodes might move to the same community, or they
might move in opposite directions (each community gains one node and
loses the other node). We use a direction indicator $D \in \{-1,1\}$
to indicate how the nodes move.  If they both move to the same
community, $D=1$; otherwise, $D=-1$.

The calculation of $\Delta^2 Q^{D,um}_t$ is given by equation 
\ref{eqn:second-derivative}. The derivation is provided in appendix
\ref{appdx:second-derivative}.

\begin{eqnarray}
\Delta^2 Q^{D,um}_t &=& 4 D \frac{(\bar{d}_u \bar{d}_m - d_u d_m)}{T^2} 
+ \frac{(w_{fs} + \bar{w}_{fs})}{T}
\label{eqn:second-derivative}
\end{eqnarray}

The computational cost of computing $\Delta^2 Q^{D,um}$ is less than
that for $\Delta Q^{D,u}$. The latter requires computing $w_{gu}$,
$w_{lu}$, $\bar{w_{gu}}$, and $\bar{w_{lu}}$, which is $O(|C_i|)$,
where $|C_i|$ represents the number of nodes in community $i$. The cost of
computing $\Delta^2 Q^{D,um}$ is reduced to $O(1)$.

\subsection{Measuring communal overlap}

In section \ref{sec:results}, we will optimize the DAMM on a given
network, recover the detected communities, $C^d$, and assess the
similarity of $C^d$ to the known communities $C^k$. For this final
step, which compares two communities, we introduce a measure that we
call \emph{communal overlap}. The statistic quantifies the similarity
between two sets of communities and is used to assess the success of
each experiment.

The foundation for communal overlap is the Jaccard index, $ J(A,B) =
\left|A \bigcap B \right|/\left|A \bigcup B \right|$, which measures
similarity
between sets, say $A$ and $B$. Each community, $C_i \in C$, is a set
of nodes. Thus, the Jaccard index provides a means for comparing two
different community configurations. Let $C^k(n)$ and $C^d(n)$
represent the known and detected communities corresponding to node
$n$. Then, if $A=C^k(n)$ and $B=C^d(n)$ the Jaccard index measures
their similarity. In this context, greater similarity implies better
detection. Communal overlap, shown in equation
\ref{eqn:communal-overlap}, computes the weighted average of the
Jaccard indices for all nodes in the network. The higher the value of
communal overlap, $\Omega \in (0,1]$, the greater the similarity
between the communal configurations $C^k$ and $C^d$, where $\Omega=1$
represents a perfect match.  $\Omega=0$ is unattainable because at the
very least, for all $n$, $C^k(n)$ and $C^d(n)$ share the node $n$ and
thus $|C^k(n) \bigcap C^d(n)| > 0$.

\begin{eqnarray}
  \label{eqn:communal-overlap}
  \Omega = \frac{1}{N} \sum_{n \in N} 
  \frac{| C^k(n) \bigcap C^d(n) |}{| C^k(n) \bigcup C^d(n) |}
\end{eqnarray}

\section{Experimental Results}
\label{sec:results}
In this section, we report on three experiments.  The first two study
stochastically generated networks with a prescribed community
structure. The third experiment involves a real world network, the
2005 National Football League (NFL) schedule.  In each case, we
measure the ability of the DAMM-enabled EO algorithm to recover known
community structure.

\subsection{Experiment I: independent contributions of positive and negative edges }
\label{sec:experiment-I}

\subsubsection{Generating  networks stochastically}
\label{sec:expI-generation}

In our tests, we generated networks with $N=64$ nodes and $|C|=4$
communities of equal size (16).
Once the communities are established, both positive and negative edges
are added to the network.
By default, positive edges are added between nodes of the same
community and negative edges are added between nodes of different
communities. An exception to this rule involves \emph{false positives}
and \emph{false negatives}, discussed in section
\ref{sec:expII-generation}.

The stochastic generation algorithm uses two parameters: the mean
number of intra-community edges per node $z_{in}$ (both nodes in the
same community), and the mean number of inter-community edges per node
$z_{out}$ (nodes in different communities). Intra-community edges are
assigned an edge weight of $e_{ij}=1$, and inter-community edges are
negatively weighted ($e_{ij}=-1$).

For any node $n \in N$ there are $|N|-1$ possible edges, discounting
self-loops. 
To generate the positively weighted edges, a pseudo-random number
$r_{in} \in [0,1)$ is generated for each potential intra-community
edge. If $r_{in} < p_{in}$ the edge is added, where $p_{in}$ is the
probability of an intra-community edge existing: $p_{in} =
z_{in}/\left(|C_i|-1\right)$.  We follow a similar procedure for
negative edges.  Each potential inter-community edge is generated with
probability $p_{out} = z_{out}/\left(|N|-|C_i|\right)$.

\subsubsection{Experimental setup}
We refer to the mean cumulative degree of a node, which combines both
the intra-community and inter-community degrees, as
$z_{cum}=z_{in}+z_{out}$. For the first experiment, we generated a
series of networks with $z_{cum}=16$. While the value of $z_{cum}$ was
held static, $z_{in} \in [0,16]$ and $z_{out} \in [0,16]$ were
dynamically adjusted and relate inversely such that $z_{out} = 16 -
z_{in}$. For each $(z_{in}, z_{out})$ parameter setting, $100$
independent networks were generated. We refer to these networks as
being dual assortative (DA). Our goal is to compare the independent
contributions of the positive and negative edges of the DA
networks. Towards this end, from each DA network, we extracted the
embedded positive assortative (PA) and negative assortative (NA)
networks. To extract the PA network, all negative edges were removed
from the original DA network. In contrast, to establish the NA
network, all positive edges were removed from the DA network. For each
network -- whether it be a DA, PA, or NA network -- the DAMM was
optimized using the EO algorithm and the community overlap $\Omega$
was assessed.

Any value of $z_{in} \ge |C_i|$ yields full intra-component
connectivity.  Thus, for $z_{in}=15$ and $z_{in}=16$, the
intra-component sub-networks are fully connected. On the other hand,
the maximum value of $z_{out}=16$ covers merely one-third of the
potential inter-component edge space.

\subsubsection{Results}
\begin{figure}
  \resizebox*{0.9\linewidth}{!}{\includegraphics[angle=-90]{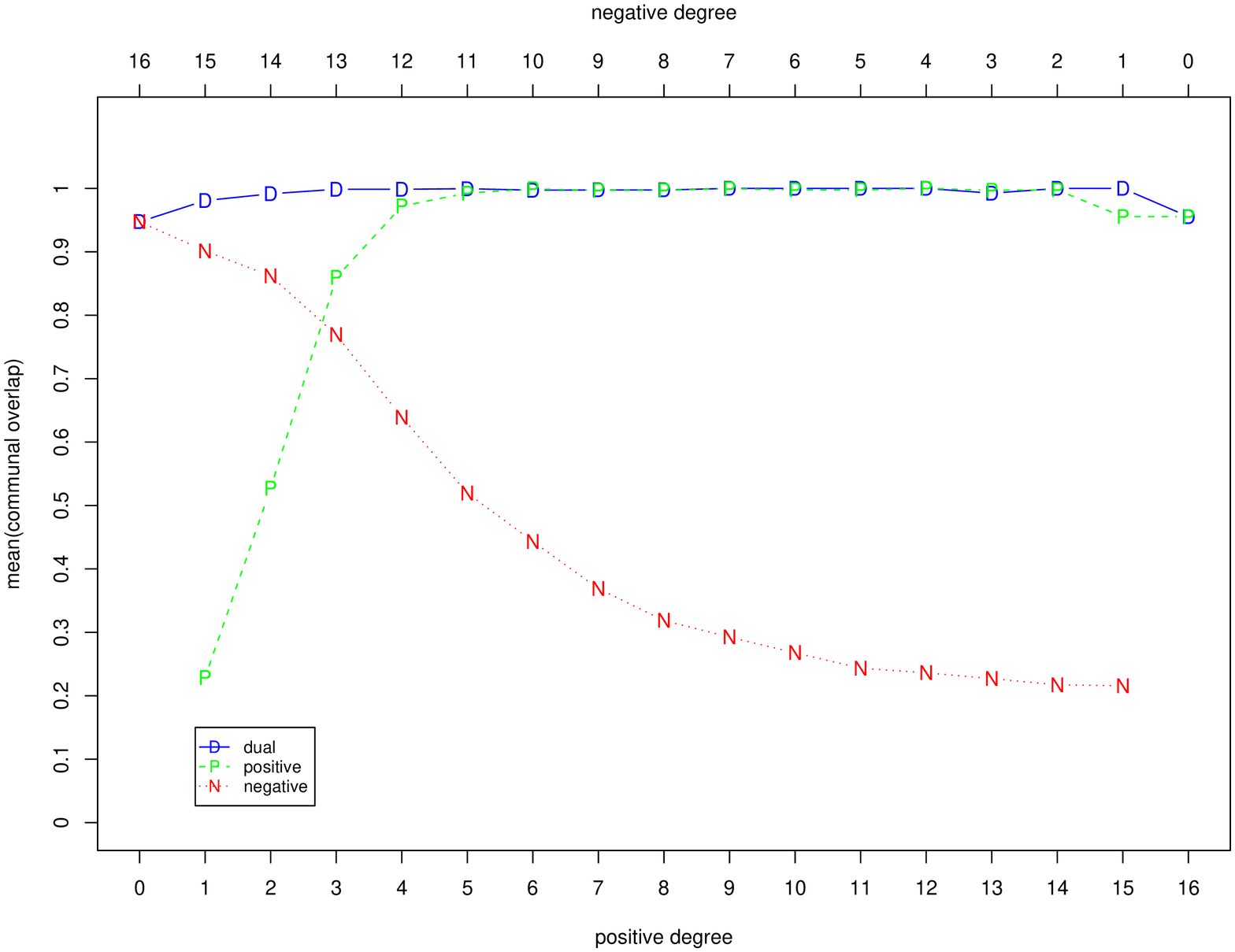}}
  \caption{Community overlap comparison for dual assortative (solid),
    positive assortative (dashed), and negative assortative (dotted)
    networks. Each data point represents the mean communal overlap for
    optimization of the DAMM on $100$ independent, computer generated
    networks.}  
  \label{fig:da-pa-na}
\end{figure}

Figure \ref{fig:da-pa-na} shows the results of the first experiment.
On the lower x-axis, the positive degree, $z_{in}$, is displayed. On
the top x-axis, the negative degree, $z_{out}$, is shown. The y-axis
represents the mean communal overlap, $\langle \Omega \rangle= 1/|G|
\sum_{i=0}^{|G|} \Omega(G_i)$, for the set of generated networks $G$
corresponding to the specified $(z_{in}, z_{out})$ setting.  The solid
curve shows results for the DA networks.  For all $(z_{in},z_{out})$
settings, $\langle \Omega_{DA} \rangle > 0.95$.  Optimization of the
DAMM on the DA networks detects the known communities with high
fidelity. The dashed curve shows DAMM-optimized PA networks. For
$z_{in} \ge 4$, the PA networks yield a $\langle \Omega_{PA} \rangle >
0.95$, which is comparable to the DA networks.  However, for $z_{in} <
4$, the community overlap values for the PA networks are significantly
less than those observed for the DA networks. This deficiency
highlights the importance of the negative edges that are removed to
create the PA networks. By removing these edges, information used by
the DAMM is lost. As a result, the detection process suffers. The
dotted curve presents the results for the NA networks. Note that for
only $z_{out}=16$ does $\langle \Omega_{NA} \rangle> 0.95$. For
$z_{out} \le 10$, $\langle \Omega_{NA} \rangle < 0.5$, which means
that, on average, there exists less than a $50\%$ overlap between the
detected and known communities.

The distance between the NA (dotted) and DA (solid) curves of figure
\ref{fig:da-pa-na} highlights the importance of the positive edges
that were removed from the original DA networks.  The mean distance
between the DA (solid) and PA (dashed) curves is $0.084$ units of
community overlap. By comparison, the mean distance between the DA and
NA curves is $0.50$ units of community overlap. Removal of the
positive edges from the DA networks for the investigated parameter
range has a significantly greater deleterious impact on community
detection.

These results provide a proof-of-principle for the DAMM.  Regardless
of the $(z_{in}, z_{out})$ setting, optimization of the DAMM yields a
high communal overlap $(\langle \Omega_{DA} \rangle > 0.95)$ on the DA
networks. When either the positive or negative edges are removed, the
detection process suffers. Note that if we simply optimize the
original modularity measure on the DA networks, the contributions of
the negative edges are ignored. By optimizing the DAMM on the PA
networks, we achieve the equivalent -- since the negative edges have
been removed, the negative information is unavailable to the DAMM.
Without the negative edges, the communal overlap yield drops. By
incorporating the contributions of both positive and negative edges,
the DAMM outperforms the original modularity measure. Furthermore, we
have demonstrated high fidelity community detection using only the
negative edges. At $(z_{in}=0, z_{out}=16)$, optimization of the DAMM
yields $\langle \Omega_{NA} \rangle > 0.95$.

\subsection{Experiment II: the impact of false positives and false
  negatives}
\label{sec:experiment-II}
The second experiment introduces ``false'' edges to the networks:
\emph{false positives} and \emph{false negatives}. A false positive is
a positively weighted edge that connects nodes in different
communities. In the language of friends and adversaries, a false
positive indicates a friendship between people of different
communities.  A false negative occurs when two nodes of the same
community are connected by a negatively weighted edge. This occurs
when there is an adversarial relationship between two people of the
same community. False positives and false negatives are routinely
found in real-world networks. Their presence contributes to the
challenge of detecting communities.

To assess the impact of the false positives and false negatives, we
generated DA networks with a fixed $(z_{in}, z_{out})$ setting, and
then exclusively added either false positives or false negatives. For
this experiment, we did not disassemble the DA networks into their PA
and NA constituents. The community detection algorithm, which optimizes
the DAMM, was applied to each DA network and the communal overlap
$\Omega$ was computed.

\subsubsection{Generating false positives and false negatives}
\label{sec:expII-generation}

To include false positives and false negatives, we introduce two
additional parameters: $f^+$, the mean number of false positives per
node, and $f^-$, the mean number of false negatives per node. To
generate false positives, we assess the unused negative edge space
following the initial edge generation phase (section
\ref{sec:expI-generation}). Assume that $E^-$ represents the entire
negative space considered for a given node in the initial phase.  We
refer to the unused subset of this space as $U^- \in E^-$ and
establish the probability $\phi^+ = f^+ / U^-$.  For each potential
edge $e_{ij} \in U^-$, we generate a random number $r^+ \in [0,1)$. If
$r^+ < \phi^+$, a positive edge between is added such that $e_{ij}=1$,
where $i$ and $j$ are known to reside in different communities. The
generation of false negatives follows a similar procedure; however,
$f^-$ dictates the likelihood of adding negatively weighted edges
between nodes residing in the same community.

\subsubsection{Experimental setup}
We generated base networks with three different settings: $(z_{in}=5,
z_{out}=16)$, $(z_{in}=7, z_{out}=16)$, and $(z_{in}=5, z_{out}=22)$.
The first parameter pair was chosen because the degree represents
one-third connectivity within both the intra-community and
inter-community subspaces.  For a given node, since $|C_i|=16$, the
maximum number of intra-component edges is $z_{in}=15$ and the maximum
number of inter-component edges is $z_{out}=48$.  Figure
\ref{fig:da-pa-na} shows that $z_{in}=5$ represents the relative
threshold for which $\Omega_{PA} > 0.95$ and $z_{out}=16$ for
$\Omega_{NA} > 0.95$. The other two parameter settings were chosen to
analyze the effect of independently increasing intra-community or
inter-community connectivity. The second pair of parameters,
$(z_{in}=7, z_{out}=16)$, was chosen to highlight the effect of
increasing $z_{in}$ when $z_{out}$ is maintained.  The increase from
$z_{in}=5$ to $z_{in}=7$ represents a $13\%$ increment in
intra-community coverage. Analogously, the third parameter pair,
$(z_{in}=5, z_{out}=22)$, represents a $13\%$ increase in the
inter-community coverage and allows us to analyze the effect of
increasing $z_{out}$ while holding $z_{in}$ steady. To these base
networks, we independently added either false positives or false
negatives. Accordingly, the edge generation parameter space is
extended to $(z_{in}, z_{out}, f^+, f^-)$ with $f^+ \in [0,8]$ and
$f^- \in [0,8]$. None of the networks contain both false positives and
false negatives: the addition of the false edges is mutually exclusive
to a single type. Thus, if $f^+>0$, then $f^-=0$; conversely, if
$f^->0$, then $f^+=0$. For each parameter setting, forty networks were
created, each with a different random number generator seed.

\subsubsection{Results}
\begin{figure}
  \resizebox{0.9\linewidth}{!}{\includegraphics[angle=-90]{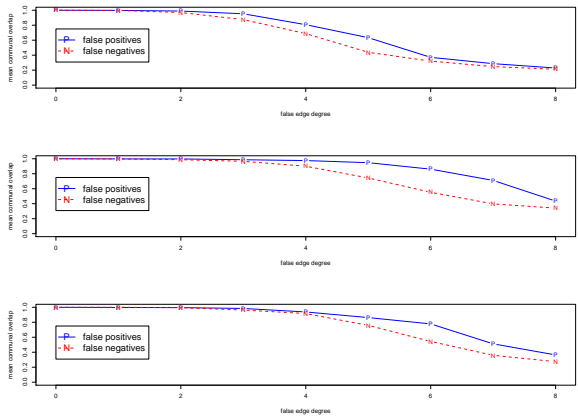}}
  \caption{The effect of false positives
    (solid) and false negatives (dashed) on communal overlap. The top graph represents base
    networks with $(z_{in}=5, z_{out}=16$); the middle graph for
    networks with $(z_{in}=7, z_{out}=16)$; and the bottom graph for
    networks with $(z_{in}=5, z_{out}=22)$. For each graph, the y-axis
    represents the mean communal overlap value for forty networks. The
    x-axis represents the mean number of either false positives or
    false negatives added to the networks.}
  \label{fig:false-pos-neg-comparison}
\end{figure}

Figure \ref{fig:false-pos-neg-comparison} shows the results on networks
with false positives and false negatives. 
In the top graph, corresponding to $(z_{in}=5, z_{out}=16, f^+,
f^-)$, we see that both curves are similar, although the
detrimental effect of the false negatives is slightly greater.  As
expected, as the rate of either false positives or false negatives
increases, $\langle \Omega \rangle$ decreases. When only a couple of
false edges are added, the known communities are detected without a
significant drop-off. For $f^+ < 3$ and $f^- < 3$, the mean communal
overlap exceeds $\langle \Omega \rangle =0.95$.  However, beyond this
range, the detection rate suffers. With reference to table
\ref{tbl:false-pos-neg}, we see that the mean communal overlap for
both curves, $M = \langle \left(\Omega^+ + \Omega^-\right)/2 \rangle$,
is $0.67$.

In the middle graph, corresponding to $(z_{in}=7, z_{out}=16, f^+,
f^-)$, and with the mean degree of intra-community edges increased
from $z_{in}=5$ to $z_{in}=7$, the effect of the additional positive
edges is observable. Note that for $f^+ < 6$ and $f^- < 4$, the mean
communal overlap $\langle \Omega \rangle > 0.95$. Thus, the range of
high fidelity detection has been extended.
Comparison to the top graph highlights another effect of the
additional information: the mean distance between the false positives
curve and the false negatives curve, denoted as $\langle \delta
\rangle$, has increased.  With reference to table
\ref{tbl:false-pos-neg}, $\langle \delta_{{z_{in}=5, z_{out}=16}}
\rangle = .058$ as compared to $\langle \delta_{z_{in}=7, z_{out}=16}
\rangle = .113$. Further, the increase in $z_{in}$ improves the
mean communal overlap for the range of both curves from $M_{z_{in}=5,
  z_{out}=16}=.67$ to $M_{z_{in}=7, z_{out}=16}=.82$.

The bottom graph presents results for $(z_{in}=5,z_{out}=22, f^+,
f^-)$, for which, in comparison to the top graph, $z_{out}$ is
increased and $z_{in}$ is unchanged. Similar to the effect
observed in the middle graph, $M$ increases in comparison to the
initial parameter setting ($M_{z_{in}=5, z_{out}=22}=.79$ as compared
to $M_{z_{in}=5, z_{out}=16}=.67$). However, unlike the case for
increasing $z_{in}$ (top graph), the mean distance between the curves,
$\langle \delta \rangle$, does not differ significantly ($\langle
\delta_{z_{in}=5, z_{out}=22}=.069 \rangle$ compared to $\langle
\delta_{z_{in}=5, z_{out}=16}=.058 \rangle$).

The second experiment establishes that the independent impact of false
positives and false negatives is influenced by the composition of the
network. An increase in either $z_{in}$ or $z_{out}$ lessens the
detrimental effect of either false positives or false negatives, as
demonstrated by the $M$ values of table
\ref{tbl:false-pos-neg}. 
Further, the additional edges (relating to $z_{in}$ and
$z_{out}$), appear to have an asymmetric effect on the impact of false
positives and false negatives. The increase in $z_{in}$ significantly
widens the gap between the false positives curve and the false
negatives curve ($\langle \delta_{z_{in}=7, z_{out}=16}
\rangle=.113$). The increase in $z_{out}$ has a much less pronounced
effect on the distance between the false positives and false negatives
curves ($\langle \delta_{z_{in}=7, z_{out}=16} \rangle$=.069).

\begin{table}[htp]
  \begin{ruledtabular}
    \caption{\label{tbl:false-pos-neg}Results for false positives and
      negatives.}
    \begin{tabular}{c|c||c|c|c|c}
      $z_{in}$&$z_{out}$&$\langle \Omega^+ \rangle$&$\langle \Omega^-
      \rangle$&$\langle \delta \rangle$&$M = \langle \frac{\Omega^+ + \Omega^-}{2} \rangle$\\
      \hline
      5&16&.70&.64&.058&.67\\
      7&16&.88&.77&.113&.82\\
      5&22&.83&.76&.069&.79\\
    \end{tabular}
  \end{ruledtabular}
\end{table}

\subsection{Experiment III: 2005 National Football League
  schedule}
\label{sec:experiment-nfl}
The third experiment uses a real dataset: the 2005 National
Football League (NFL) schedule. From the dataset, we construct networks
representing the correlation of team schedules. Each team is
represented by a node. Edges between nodes are weighted to indicate
the correlation of the two team's schedules. Teams that play similar
opponents show positive correlations. Teams that play dissimilar
schedules are negatively correlated. The network contains both
positively and negatively weighted edges and is thus dual assortative.
Because it is possible to compute the correlation between any two team
schedules, the network is fully connected. However, the objective of
the experiment is to examine the efficacy of optimizing the DAMM on
partial representations of the dual assortative network. Accordingly,
only a subset of the possible edges are represented in any given
generated network.

The NFL consists of thirty-two teams split into two conferences.
Within each conference, teams are grouped into four divisions of four
teams apiece. Each team plays sixteen games.  Six of these games are
the result of a team playing its three division rivals twice each. In
addition, each division is paired with one division of the same
conference and a second division that resides in the other conference.
For each team, these division-versus-division games account for eight
additional games (bringing the running tally to fourteen games). The
final two opponents for each team result from games against teams from
the same conference, but not involved in the division-versus-division
matchup.  In total, each team faces thirteen unique opponents.

Through extensive analysis using the EO algorithm, we identified four
optimal and two near-optimal communal alignments for the fully
connected NFL schedule correlation network. We refer to these communal
alignments as the \emph{known} optimal configurations. The four
optimal alignments each consist of three communities (one community
consisting of $8$ teams and the other two communities containing $12$
teams apiece). In each case, the $8$ team community is comprised of
two divisions from the same conference that are pitted in a
division-versus-division matchup. Each of the $12$ team communities
contain three divisions, with one division being involved in an
intra-conference division-versus-division matchup with one of the
other divisions and an inter-conference division-versus-division
matchup with the remaining division. Both of the near-optimal communal
alignments consist of four communities. In one case, each community
contains two divisions pitted in an inter-conference
division-versus-division matchup. In the other, each community
consists of two divisions pitted in an intra-conference
division-versus-division matchup.

With regards to the four optimal communal alignments, the NFL schedule
correlation network contains both \emph{false positives} and
\emph{false negatives}. In each alignment, there exist teams sharing
positively weighted edges that belong to different communities. These
edges constitute the false positives. Further, in each alignment,
there exist teams within the same community that share negatively
weighted edges. These edges constitute false negatives. 

\subsubsection{Generation of networks}
To study the performance of the DAMM on the NFL network, we first
optimized it on various \emph{partial} representations of the
NFL schedule correlation network. We then compared the detected
communal alignment to the set of known optimal configurations and
identified the closest match.  The best communal overlap score from
this series of comparisons was recorded as the communal overlap value.

The fully connected NFL schedule correlation network contains $992$
edges (discounting self-loops). Of these edges, $352$ ($35$ percent)
are positively weighted and $640$ are negatively weighted.  We
generated the partial representation using a procedure similar to the
edge generation algorithm used to generate the partial representations
described in section \ref{sec:expI-generation}.  Instead of computing
probability thresholds (such as $p_{in}$ and $p_{out}$) from a mean
degree (such as $z_{in}$ and $z_{out}$), we simply used the
probability thresholds as parameters.  Each possible positively
weighted edge of the full representation was selected with probability
$p^{+}$ and each negative edge was selected with probability $p^{-} =
1 - p^{+}$.

The stochasticity of this process guarantees that with high
probability individual nodes of a partial representation will have
varying degree.  Because of this asymmetry, certain nodes are more
difficult to classify than others. These asymmetries could yield
situations where the optimal experimental DAMM configuration will not
concur with the known optimal configurations. In such a case, a
sub-optimal communal overlap will result.

Our goal is to examine whether, on average, the DA information
utilized by the DAMM leads to better community detection relative to
either the PA or NA information in isolation. Recall that we create
the PA network by removing all negative edges from the corresponding
DA network; whereas, for the NA network, we remove all positive edges
from the DA network.

\subsubsection{Experimental setup}
We explored the parameter range $p^{+} \in [0,1]$ and $p^{-} \in
[0,1]$. 
All of the studied networks were partial representations of the fully
connected network.  For each $(p^{+}, p^{-})$ setting, we generated
$40$ networks. Similar to the experiment of section
\ref{sec:experiment-I}, in each case, we separately optimized the DAMM
on the DA network, the PA network, and the NA network.

\subsubsection{Results}
\begin{figure}
  \resizebox{0.9\linewidth}{!}{\includegraphics[angle=-90]{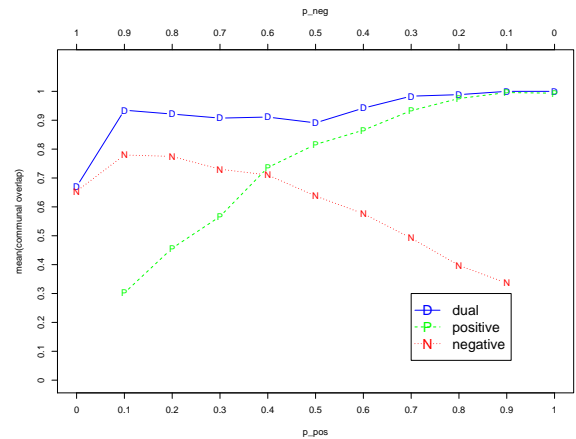}}
  \caption{Community overlap measures for the 2005 NFL schedule
    network. The bottom x-axis represents $p^{+}$ and the top x-axis
    represents $p^{-}$. The y-axis represents the mean communal
    overlap, $\langle \Omega \rangle$. The different curves present
    information regarding the different types of networks upon which
    the DAMM is optimized: DA networks (solid), PA networks (dashed),
    and NA networks (dotted). Optimization of the DAMM on the DA networks
    yields as good or better mean communal overlap values than for
    either the PA or NA networks. By utilizing both the positively and
    negatively weighted edges, optimization of the DAMM provides
    better community detection than the original modularity measure
    that operates only on positively weighted edges.}
  \label{fig:nfl-2005-comm-overlap}
\end{figure}

Figure \ref{fig:nfl-2005-comm-overlap} gives the results of the third
experiment. The bottom x-axis represents the $p^{+}$ values; the top
x-axis represents the $p^{-}$ values. The y-axis represents the mean
communal overlap, $\langle \Omega \rangle$, for the corresponding
($p^{+}, p^{-})$ thresholds. Each data point represents the mean
communal overlap resulting from optimization of the DAMM on $40$
independent, randomly generated partial networks with the same
prescribed thresholds.

For each $(p^{+}, p^{-})$ setting, optimization on the DA networks
yields an equal or higher $\langle \Omega \rangle$ value than for
either the PA or NA networks. Only at $(p^{+}=1, p^{-}=0)$ and
$(p^{+}=0, p^{-}=1)$ do $\langle \Omega_{PA} \rangle = \langle
\Omega_{DA} \rangle$ and $\langle \Omega_{NA} \rangle = \langle
\Omega_{DA} \rangle$, respectively. At these settings there are either
exclusively positive or exclusively negative edges, and thus, these DA
networks are equivalent to the respective PA or NA cases. For all
parameter settings at which there are both positive and negative
edges, the DAMM uses both types of information and achieves higher
mean communal overlap values. Despite the presence of $1.8$ times more
negative edges than positive edges, the positive edges provide more
information for community detection.  Using only negative edges, at
$(p^{+}=0, p_-=1)$, optimization of the DAMM detects a sub-optimal
communal alignment. On the other hand, using only positive edges (at
$(p^{+}=1, p_-=0)$) optimization yields an optimal mean communal
overlap value.  Figure \ref{fig:nfl-2005-comm-overlap} highlights the
asymmetry regarding the amount of information provided by the positive
edges as compared to the negative edges. The mean distance between the
DA and PA curves is $0.20$ communal overlap units; whereas, the mean
distance between the DA and NA curves is $0.34$ communal overlap
units. The positive edges contribute more to the community detection
process.  As expected, as $p^{+}$ increases $\langle \Omega_{PA}
\rangle$ increases. Similarly, as $p^{-}$ increases, $\langle
\Omega_{NA} \rangle$ increases.

\section{Summary and conclusions}
\label{sec:conclusion} 
The DAMM provides a way to assess community structure in networks
containing both positively and negatively weighted edges. This extends
the paradigm of the friendship network to that of a \emph{friends and
adversaries} network.  Negative information, previously ignored, now
provides useful, additional information to community detection
algorithms.

The efficacy of the DAMM was demonstrated, both for stochastically
generated synthetic networks and a real-world example based on the
2005 NFL schedule.  Furthermore, the experiments revealed the
asymmetry in the information provided by positive and negative
edges. This asymmetry is due to the greater specificity provided by a
positive edge given that more than two communities exist.

The contributions of the DAMM are two-fold. First, we can now analyze
networks containing solely negative information. Second, the DAMM
improves community detection in networks containing both positive and
negative information. An example of such a network is one in which
edge weights are based on a similarity metric, such as correlation,
that can assume either positive or negative values. The NFL schedule
correlation network of section \ref{sec:experiment-nfl} provides a
real-world example. The DAMM expands the domain of problems for which
community detection algorithms can be applied.

\appendix
\section{Derivation of $\Delta Q^{D,u}$}
\label{appdx:first-derivative}

As expressed in equation \ref{eqn:mod-dual}, the DAMM, $Q^D$, involves
summing the independent contributions of all communities $i$. We can
represent the contribution a given community $i$ with a local DAMM
value, denoted as $q_i$, such that $Q^D =\sum_i q_i$. Further, if we
wish to independently assess the positive and negative edge weight
contributions of each given community, we can write $Q^D = \sum_i
q_i^+ + q_i^-$.

Computing $\Delta Q^{D,u}$ entails comparing the DAMM value from
timestep $t$ to the DAMM value at time $t+1$ that results from
migrating node $u$. We denote the DAMM value at time $t$ as $Q^{D}_t$
and the DAMM value following the migration of node $u$ as
$Q^{D}_{t+1(u)}$ Accordingly, $\Delta Q^{D,u}_t$ can be expressed as:

\begin{eqnarray}
\Delta Q^{D,u}_t &=& Q^{D}_{t+1(u)} - Q^{D}_t
  \label{eqn:qDamm1}
\end{eqnarray}.

Utilizing the local DAMM notation, we can rewrite equation 
\ref{eqn:qDamm1} as:

\begin{eqnarray}
\Delta Q^{D,u}_t &=& (\sum_i q^{+}_{i,t+1 (u)} + q^{-}_{i,t+1 (u)}) -
(\sum_i q^{+}_{i,t} + q^{-}_{i,t})
\label{eqn:qDamm2}
\end{eqnarray}

, where $q^+_{i,t}$ represents the positive edge contribution of the
local DAMM value for community $i$ at time $t$ (before migrating the
node $u$) and $q^+_{i,t+1 (u)}$ represents the positive edge
contribution of the same community following the migration of node
$u$.  The subscript $t+1 (u)$ is used to indicate that a node $u$ has
been migrated.

By grouping the positive local DAMM values and the negative local
DAMM values separately, we can write: 

\begin{eqnarray}
\Delta Q^{D,u}_t &=& \sum_i (q^{+}_{i,t+1 (u)} - q^+_{i,t}) + \sum_i
(q^{-}_{i,t+1 (u)} - q^-_{i,t})\\ 
&=& \Delta Q^{+,u}_t + \Delta Q^{-,u}_t
\label{eqn:qDamm3}
\end{eqnarray}.

By moving a single vertex from one community to another, as is done
during the division process, only two communities are affected.
All other communities remain unchanged. One community gains a new
node. We refer to this community as the \emph{gain} community, and
denote it as $C_g$. Conversely, the other affected community loses a
node. We denote this \emph{loss} community as $C_l$. Since only the
$C_g$ and $C_l$ communities are affected by a vertex move, the only
local DAMM values that change are those relating to these communities
-- $q_g$ and $q_l$.  The local DAMM contributions of all other
communities, $\{q_i| i \ne g, i \ne l\}$, remain unchanged. Thus, we
need only to assess the change in DAMM for the gain and loss
communities. Using this information, we rewrite $\Delta Q^{+,u}_t$ as:

\begin{eqnarray}
\Delta Q^{+,u}_t &=& \sum_i q^{+}_{i,t+1 (u)} - q^+_{i,t}\\
&=& (q^{+}_{g,t+1 (u)} - q^+_{g,t}) + (q^{+}_{l,t+1 (u)} - q^+_{l,t})\\
&=& \Delta q^{+,u}_{g,t} + \Delta q^{+,u}_{l,t}
\end{eqnarray}

, where $q^{+}_{g,t}$ denotes the local DAMM value for the positive
edges associated with the gain community at time $t$, $q^+_{l,t}$
denotes the local DAMM value for the positive edges of the loss
community at time $t$, and $q^{+}_{g,t+1 (u)}$ represents the local
DAMM value for the positive edge contributions of the gain community
following the migration of node $u$.

We can now separately analyze the contributions of $\Delta
q^{+,u}_{g,t}$ and $\Delta q^{+,u}_{l,t}$ and reassemble the terms to
establish $\Delta Q^{+,u}_t$. We denote the node to be moved as $u$
and introduce the following notation: $\{w_{gg} = \sum_{rs} e_{rs} | r
\in C_g, s \in C_g\}$ to represent the cumulative intra-community
positive edge weight for the gain community, $\{w_{gu} = \sum_{rs}
e_{rs} | r \in C_g, s = u\}$ to represent the cumulative edge weight
of the gain community connected to node $u$, and $d_u$ to represent
the positive degree of node $u$.  Note that $q^+_{g,t}$, which
represents the contribution of the positive edges in the gain
community prior to moving the node, is defined as:

\begin{eqnarray}
q^+_{g,t} &=& \frac{2w_{gg}}{T} - \left(\frac{a_g}{T}\right)^2
\label{eqn:pos-gain-pre}
\end{eqnarray}.

We use the notation $q^{+}_{g,t+1 (u)}$ to denote the contribution of
the gain community positive edges after migrating node $u$. Following
the migration, the gain community contains an additional node.
Accordingly, both the cumulative intra-community positive edge weight,
$w_{gg}$, and the total cumulative positive edge weight of the gain
community, $a_g$, are subject to change. More specifically, the
intra-community positive edge weight is updated as $w_{gg,t+1(u)} =
w_{gg,t} + w_{gu,t}$, where $w_{gu,t}$ represents the cumulative
positive edge weight connecting the node $u$ to the gain community
prior to the migration. Furthermore, the cumulative positive edge
weight of the gain community is updated as $a_{g,t+1(u)} = a_{g,t} +
d_u$. Using these updates, we define $q^{+}_{g,t+1 (u)}$ as:

\begin{eqnarray}
  q^{+}_{g,t+1 (u)} &=& \frac{2(w_{gg} + w_{gu})}{T} - \left(\frac{a_g
      + d_u}{T}\right)^2
\label{eqn:pos-gain-post}
\end{eqnarray}.

By subtracting equation \ref{eqn:pos-gain-pre} from equation
\ref{eqn:pos-gain-post}, we establish $\Delta q^{+,u}_{g,t}$ as provided in
equation \ref{eqn:delta-pos-gain}:

\begin{eqnarray}
  \Delta q^{+,u}_{g,t} &=& q^{+}_{g,t+1 (u)} - q^+_{g,t}\\
  &=& \frac{2w_{gu}}{T} - \frac{2d_u}{T^2} \left(a_g +
    \frac{d_u}{2}\right) 
  \label{eqn:delta-pos-gain}
\end{eqnarray}

Similarly, the positive edge contribution of the loss group, denoted
as $\Delta q^{+,u}_{l,t}$, is defined by equation
\ref{eqn:delta-pos-loss}.

\begin{eqnarray}
  \Delta q^{+,u}_{l,t} &=& q^{+}_{l,t+1 (u)} - q^{+}_{l,t}\\
  &=& \left[\frac{2(w_{ll} - w_{lu})}{T} -
    \left(\frac{a_l-d_u}{T}\right)^2 \right] \nonumber\\
  &&- \left[\frac{2w_{ll}}{T} - \left(\frac{a_l}{T}\right)^2 \right]\\
  &=& \frac{-2w_{lu}}{T} + \frac{2d_u}{T^2}\left(a_l -
    \frac{d_u}{2}\right)
  \label{eqn:delta-pos-loss}
\end{eqnarray}

By assembling equations \ref{eqn:delta-pos-gain} and
\ref{eqn:delta-pos-loss}, we establish $\Delta Q^{+,u}_t$ as provided by
equation \ref{eqn:delta-pos}.

\begin{eqnarray}
  \Delta Q^{+,u}_t &=& \Delta q^{+,u}_{g,t} + \Delta q^{+,u}_{l,t}\\
  &=& 2 \left[\left(\frac{w_{gu}-w_{lu}}{T}\right) + \frac{d_u}{T^2}
    \left(a_l-a_g-d_u)\right) \right]
  \label{eqn:delta-pos}
\end{eqnarray}

Following a similar logic, we establish $\Delta Q^{-,u}_t$. For
brevity, we provide the result in equation \ref{eqn:delta-neg}, where
$\bar{w}_{gu}$ represents the cumulative negative edge weight of the
gain community connected to node $u$ and $\bar{d}_u$ represents the
negative degree of the node $u$.

\begin{eqnarray}
  \Delta Q^{-,u}_t &=& \Delta q^{-,u}_{g,t} + \Delta q^{-,u}_{l,t}\\
&=& 2 \left[\left(\frac{\bar{w}_{lu}-\bar{w}_{gu}}{T}\right) +
    \frac{\bar{d}_u}{T^2} \left(\bar{a}_l-\bar{a}_g-\bar{d}_u)\right) \right]
  \label{eqn:delta-neg}
\end{eqnarray}

Finally, we establish $\Delta Q^{D,u}_t$ by assembling equations
\ref{eqn:delta-pos} and \ref{eqn:delta-neg}: 

\begin{eqnarray}
  \Delta Q^{D,u}_t &=& \Delta Q^{+,u}_t + \Delta Q^{-,u}_t\\
  &=& 2 \left[\left(\frac{w_{gu}-w_{lu}}{T}\right) + \frac{d_u}{T^2}
    \left(a_l-a_g-d_u)\right) \right] \nonumber \\
  && + 2 \left[\left(\frac{\bar{w}_{lu}-\bar{w}_{gu}}{T}\right)
   + \frac{\bar{d}_u}{T^2}
   \left(\bar{a}_g-\bar{a}_l+\bar{d}_u)\right) \right]
  \label{eqn:qDamm-appdx}
\end{eqnarray}

\section{Derivation of $\Delta^2 Q^{D,um}$}
\label{appdx:second-derivative}
As a first step of our derivation, we concentrate solely on the
contributions of the positive edges.  First, we analyze the difference
$\Delta^2 Q^{+,um}_t = \Delta Q^{+,u}_{t+1 (m)} - \Delta Q^{+,u}_{t}$,
where $\Delta Q^{+,u}_{t+1 (m)}$ represents the change in DAMM that
would result from the migration of node $u$ if node $m$ were already
to have been migrated given the current configuration. $\Delta
Q^{+,u}_t$, which pertains exclusively to positive edges, was provided
in equation \ref{eqn:delta-pos}. To simplify the derivation, we
independently assess the first and second terms of equation
\ref{eqn:delta-pos}, such that $A^u_t=\frac{w_{gu}-w_{lu}}{T}$ and
$B^u_t = (a_l-a_g-d_u)$.

After migrating node $m$, the value of $A$ may be altered -- and, this
change should be reflected in $A^u_{t+1 (m)}$. Note the two terms
involved in $A^u_t$: $w_{gu}$ and $w_{lu}$. Each term measures the
cumulative positive edge weight connecting a community -- either the
gain or loss community -- to the node $u$ prior to the migration of
node $m$. The migration of node $m$ may or may not alter the
cumulative positive edge weight between each community and node $u$.
If node $m$ was migrated to the community currently not occupied by
$u$, the \emph{gain} community, $w_{gu}$ will increase such that
$w_{gu,t+1} = w_{gu,t} + e_{mu}$, where $e_{mu}$ represents the
positive edge weight between the $m$ and $u$ nodes.  Otherwise, if the
$m$ node moved in the opposite direction, $w_{gu,t+1 (m)} = w_{gu,t} -
e_{mu}$.  Using the direction indicator $D$, we can write $w_{gu,t+1
  (m)} = w_{gu,t} + De_{mu}$. Similarly, we can write $w_{lu,t+1 (m)}
= w_{lu,t} - De_{mu}$.  By assembling the two terms, we express
$A^u_{t+1 (m)}$ as:

\begin{eqnarray}
A^u_{t+1 (m)} &=& \frac{(w_{gu} + De_{mu}) - (w_{lu} - De_{mu})}{T}
\end{eqnarray}.

By subtracting $A^u_t$ from $A^u_{t+1 (m)}$, we establish $\Delta
A^u_t$ as shown in equation \ref{eqn:delta-A}.

\begin{eqnarray}
\Delta A^u_t &=& A^u_{t+1 (m)} - A^u_t\\
&=& \frac{(w_{gu} + De_{mu}) - (w_{lu} - De_{mu})}{T} \nonumber\\
 && - \frac{(w_{gu} - w_{lu})}{T}\\
&=& \frac{2 D e_{mu}}{T}
\label{eqn:delta-A}
\end{eqnarray}.

The first two terms of $B$, $a_l$ and $a_g$, are similarly affected by
the migration of node $m$. These terms represent the cumulative
positive edge weight of the loss and gain communities,
respectively. The updated terms can be expressed as $a_{l,t+1 (m)} =
a_{l,t} - Dd_m$ and $a_{g,t+1 (m)} = a_{g,t} + Dd_m$, where $d_m$
represents the positive degree of the $m$ node.  Accordingly, we can
write $B^u_{t+1 (m)}$ as:

\begin{eqnarray}
B^u_{t+1 (m)} &=& (a_l - Dd_m) - (a_g + D d_m) - d_u
\end{eqnarray}.

By subtracting $B^u_t$ from $B^u_{t+1 (m)}$, we establish $\Delta B^u_t$
as seen in equation \ref{eqn:delta-B}.

\begin{eqnarray}
\Delta B^u_t &=& B^u_{t+1 (m)} - B^u_t\\
 &=&  \left[(a_l - D d_m) - (a_g + D d_m) - d_u \right]
\nonumber\\
&& - (a_l-a_g-d_u)\\
&=& -2 D d_m
\label{eqn:delta-B}
\end{eqnarray}

We now utilize $\Delta A^u_t$ and $\Delta B^u_t$ to establish the positive edge
contribution to $\Delta^2 Q^{D,um}$:

\begin{eqnarray}
\Delta^2 Q^{+,um}_t &=& 2 \left[ \Delta A^u_t + \frac{d_u}{T^2} \Delta B^u_t
\right]\\
&=& 2 \left[ \frac{2De_{mu}}{T} + \frac{d_u}{T^2} (-2Dd_m) \right]\\
&=& 4D \left[ \frac{e_{mu}}{T} - \frac{d_u d_m}{T^2} \right]
\label{eqn:second-derivative-pos}
\end{eqnarray}.

Following a similar approach, it is possible to establish the negative
edge contribution to $\Delta^2 Q^{D,um}$:

\begin{eqnarray}
  \Delta^2 Q^{-,um}_t &=& 4D \left[ \frac{\bar{d}_u \bar{d}_m}{T^2} +
    \frac{\bar{e}_{mu}}{T}\right]
\label{eqn:second-derivative-neg}
\end{eqnarray}.

By assembling $\Delta^2 Q^{+,um}_t$ of equation \ref{eqn:second-derivative-pos}
and $\Delta^2 Q^{-,um}_t$ of equation \ref{eqn:second-derivative-neg}, we
establish the generalized equation:

\begin{eqnarray}
\Delta^2 Q^{D,um}_t &=& \Delta^2 Q^{+,um}_t + \Delta^2 Q^{-,um}_t\\
&=& 4 D \left[ \frac{(\bar{d}_u \bar{d}_m - d_u d_m)}{T^2} 
+ \frac{(e_{mu} + \bar{e}_{mu})}{T} \right]
\end{eqnarray}.

\begin{acknowledgments}
  The authors would like to thank Cristopher Moore and Petter Holme
  for discussions regarding material in this paper as well as the
  University of New Mexico Center for High Performance Computing for
  donating computational cycles to conduct the experiments.
\end{acknowledgments}

\bibliography{dual-mod}

\end{document}